\RequirePackage{fix-cm}
\documentclass[twocolumn,epjc3]{svjour3}  
\smartqed  
\RequirePackage{graphicx}
\usepackage{graphicx}
\usepackage{subcaption}
\usepackage{xcolor}

\usepackage[english]{babel}
\RequirePackage{latexsym}
\usepackage{amsmath}
\usepackage{amssymb}
\RequirePackage[numbers,sort&compress]{natbib}
\usepackage[
  colorlinks=true,
  linkcolor=blue,
  citecolor=magenta,
  urlcolor=blue
]{hyperref}
\usepackage{hyphenat}
\usepackage[displaymath, mathlines]{lineno}
\journalname{Eur. Phys. J. C}
\begin{document}

\title{Probing dark matter interactions with a RES-NOVA prototype cryogenic detector}

\author{
The \mbox{\protect{\sc{RES--NOVA}}} collaboration\thanksref{a}
        \and  \\[2mm]
D.~Alloni\thanksref{LENA, INFN-PV}
\and
G.~Benato\thanksref{GSSI, LNGS}
\and
P.~Carniti\thanksref{UNIMIB, INFN-MIB}
\and
M.~Cataldo\thanksref{UNIMIB, INFN-MIB}
\and
L.~Chen\thanksref{SICCAS}
\and
M.~Clemenza\thanksref{UNIMIB, INFN-MIB}
\and
M.~Consonni\thanksref{UNIMIB, INFN-MIB}
\and
G.~Croci\thanksref{UNIMIB, INFN-MIB}
\and
I.~Dafinei\thanksref{Roma1}
\and
F.A.~Danevich\thanksref{GSSI, INR, CZ}
\and
C.~de~Vecchi\thanksref{INFN-PV}
\and
D.~Di~Martino\thanksref{UNIMIB, INFN-MIB}
\and
E.~Di~Stefano\thanksref{INFN-MIB, DISAT}
\and
N.~Ferreiro Iachellini\thanksref{UNIMIB, INFN-MIB,b}
\and
F.~Ferroni\thanksref{GSSI,Roma1}
\and
F.~Filippini\thanksref{UNIMIB, INFN-MIB, DISAT}
\and
S.~Ghislandi\thanksref{MIT}
\and
A.~Giachero\thanksref{UNIMIB, INFN-MIB}
\and
L.~Gironi\thanksref{UNIMIB, INFN-MIB}
\and
C.~Gotti\thanksref{INFN-MIB}
\and
D.L.~Helis\thanksref{LNGS}
\and
D.V.~Kasperovych\thanksref{INR}
\and
V.V.~Kobychev\thanksref{INR}
\and
G.~Marcucci\thanksref{UNIMIB, INFN-MIB}
\and
A.~Melchiorre\thanksref{LNGS, UNIVAQ,c}
\and
A.~Menegolli\thanksref{INFN-PV, PV-FIS}
\and
S.~Nisi\thanksref{LNGS}
\and
M.~Musa\thanksref{INFN-PV, PV}
\and
L.~Pagnanini\thanksref{GSSI,LNGS}
\and
L.~Pattavina\thanksref{UNIMIB, INFN-MIB}
\and
G.~Pessina\thanksref{INFN-MIB}
\and
S.~Pirro\thanksref{LNGS}
\and
S.~Pozzi\thanksref{INFN-MIB}
\and
M.C.~Prata\thanksref{INFN-PV}
\and
A.~Puiu\thanksref{LNGS}
\and
S.~Quitadamo\thanksref{UNIMIB, INFN-MIB}
\and
M.P.~Riccardi\thanksref{INFN-PV, PV}
\and
M.~Ricci\thanksref{Roma1}
\and
M.~Rossella\thanksref{INFN-PV}
\and
R.~Rossini\thanksref{INFN-PV, PV-FIS}
\and
E.~Sala\thanksref{INFN-MIB, IBS}
\and
F.~Saliu\thanksref{INFN-MIB, DISAT}
\and
A.~Salvini\thanksref{LENA, INFN-PV}
\and
V.I.~Tretyak\thanksref{LNGS, INR, CZ}
\and
L.~Trombetta\thanksref{UNIMIB, INFN-MIB}
\and
D.~Trotta\thanksref{UNIMIB, INFN-MIB}
\and
H.~Yuan\thanksref{SICCAS}
}

\thankstext{a}{e-mail: res-nova@unimib.it}
\thankstext{b}{e-mail: nahuel.ferreiroiachellini@unimib.it}
\thankstext{c}{e-mail: andrea.melchiorre@graduate.univaq.it}


\institute{Laboratorio Energia Nucleare Applicata, Via Aselli 41, I-27100 Pavia, Italy \label{LENA}
\and
INFN Sezione di Pavia, Via Bassi 6,  I-27100 Pavia, Italy \label{INFN-PV}
\and
Gran Sasso Science Institute,  Viale F. Crispi 7, I-67100 L’Aquila, Italy \label{GSSI}
\and
INFN Laboratori Nazionali del Gran Sasso, Via G. Acitelli 22, I-67100 Assergi, Italy \label{LNGS}
\and
Dipartimento di Fisica, Universit\`a di Milano - Bicocca, Piazza della Scienza 3, I-20126 Milano, Italy \label{UNIMIB}
\and
INFN Sezione di Milano - Bicocca, Piazza della Scienza 3, I-20126 Milano, Italy \label{INFN-MIB}
\and
Shanghai Institute of Ceramics, CAS, 1295 Dingxi Road, Shanghai 200050, P.R. China \label{SICCAS}
\and
INFN Sezione di Roma-1, P.le Aldo Moro 2, I-00185 Roma, Italy \label{Roma1}
\and
Institute for Nuclear Research of NASU, 03028 Kyiv, Ukraine \label{INR}
\and
Institute of Experimental and Applied Physics, Czech Technical University in Prague, Husova 240/5, 110 00 Prague 1, Czech Republic \label{CZ}
\and
DISAT, Universit\`a di Milano - Bicocca, Piazza della Scienza 1, I-20126 Milano, Italy \label{DISAT}
\and
Massachusetts Institute of Technology, Cambridge, MA 02139, USA\label{MIT}
\and
Dipartimento di Scienze Fisiche e Chimiche, Università degli Studi dell’Aquila, I-67100 L’Aquila,
Italy \label{UNIVAQ}
\and
Dipartimento di Fisica, Universit\`a di Pavia, Via Bassi 6,  I-27100 Pavia, Italy \label{PV-FIS}
\and
Dipartimento di Scienze della Terra e dell'Ambiente, Universit\`a di Pavia, Via Ferrata 7,  I-27100 Pavia, Italy \label{PV}
\and
Center for Underground Physics, Institute for Basic Science, 34126 Daejeon, Korea \label{IBS}
}

\date{Received: date / Accepted: date}

\maketitle

\begin{abstract}
We report on the operation of a 13~g PbWO$_4$ crystal, grown from archaeological Pb and operated as a cryogenic calorimeter in an underground environment. Read out with a Ge thermistor, the detector achieves a low energy threshold and, for the first time, enables the derivation of a dark matter exclusion limit using PbWO$_4$ as target material, for both spin--dependent interactions on neutrons and spin--independent interactions.
Although limited in mass and not representative of the final RES--NOVA detector design, this prototype demonstrates effective control of mechanical vibrations and low--energy noise in a cryogenic system, which is a key requirement for rare--event searches. The experiment therefore provides a proof of principle for the RES--NOVA detection concept, validating the use of archaeological Pb--based PbWO$_4$ crystals, low--background operation, and robust data--analysis procedures. These results establish a solid technological and methodological foundation for future RES--NOVA detectors employing larger target masses and advanced thermal readout technologies.

\keywords{Dark Matter \and Cryogenic Detector \and Archaeological Pb}
\end{abstract}

\section{Introduction}
\label{intro}

The existence of dark matter (DM) is a well-established paradigm in modern cosmology and astrophysics, supported by a wide range of observations, from galactic rotation curves to the anisotropies of the cosmic microwave background and the formation of large-scale structure. Yet, despite its dominant role in the matter content of the Universe, the particle nature of DM remains elusive~\cite{Billard_2022}.

In recent years, the field of direct DM detection has undergone significant advancements in sensitivity. However, no unambiguous signal has yet been observed. As the experimental landscape approaches the so-called \emph{neutrino fog}~\cite{Fog}, where Coherent Elastic Neutrino-Nucleus Scattering (CE$\nu$NS)~\cite{Freedman:1973yd} becomes a dominant background, the need for diversified detection strategies becomes increasingly evident.

In this context, a multi-target and multi-detector approach is not only desirable but essential. Exploring different nuclear targets offers a powerful handle to disentangle potential DM signals from backgrounds, given the different dependencies of the interaction rates on nuclear parameters such as mass number and spin content~\cite{Billard_2022}. Furthermore, leveraging different detector technologies helps mitigate systematic uncertainties and enhances the robustness of any potential discovery.

Interestingly, the quest for direct DM detection shares deep experimental and conceptual similarities with the detection of low-energy neutrinos via CE$\nu$NS. Both processes involve low-energy nuclear recoils and require detectors with low energy thresholds, ultra-low background environments, and high radiopurity materials. As such, the development of detectors optimized for CE$\nu$NS can directly benefit DM searches, and vice versa~\cite{Billard_2022}.

In this paper, we report on a novel direct dark matter (DM) search that, for the first time, employs a PbWO$_4$ cryogenic detector fabricated from archaeological Pb~\cite{Pattavina:2019pxw} and operated by the RES--NOVA collaboration. The RES--NOVA experiment is currently under commissioning and is primarily dedicated to the detection of low--energy neutrinos from astrophysical sources via coherent elastic neutrino--nucleus scattering~\cite{Pattavina:2020cqc}. Thanks to its old age, the archaeological Pb used in this work is expected to exhibit an exceptionally low intrinsic radioactive background, making it particularly well suited for rare--event searches~\cite{Pattavina:2020ota,Pattavina:2020cqc,Beeman:2012wz}. 

The results presented here represent an important step toward establishing PbWO$_4$ as a competitive target material for both DM and coherent elastic neutrino--nucleus scattering (CE$\nu$NS) detection, exploiting its high mass number and favorable nuclear properties.

\section{Archaeological Pb as a target material for elusive-rate searches}

Direct searches for DM and the detection of low--energy neutrinos via CE$\nu$NS involve extremely low interaction rates, due to the small cross sections of these processes. Typical CE$\nu$NS cross sections lie in the range $10^{-39}$--$10^{-41}~\text{cm}^2$~\cite{Cadeddu:2023tkp}, while for DM--nucleus elastic scattering current exclusion limits probe cross sections below $10^{-46}~\text{cm}^2$~\cite{Billard_2022} for DM masses in the GeV--TeV range. Consequently, the detection of such rare events requires detectors with exceptionally low background levels and very low energy thresholds.

To achieve this goal, particular care must be devoted to the selection and preparation of detector materials. One of the dominant background sources in rare-event searches is the intrinsic radioactivity of the materials used in the detector construction and its surrounding infrastructure~\cite{Laubenstein:2020rbe}. Long-lived isotopes from the uranium and thorium decay chains, as well as $^{210}$Pb and cosmogenic activation products, can generate signals in the energy region of interest and thus mimic or obscure the expected signal~\cite{Clemenza:2011zz}.

In this context, archaeological Pb offers a unique advantage thanks to its extremely long cool-down time since production. Having been shielded from cosmic rays for centuries, typically underwater or underground, this material exhibits strongly suppressed levels of radioactive isotopes, in particular $^{210}$Pb and its progeny~\cite{Pattavina:2019pxw}. Its intrinsic radiopurity makes archaeological Pb an excellent candidate for use in low--background detectors, either as passive shielding or as an active component, as in the case of PbWO$_4$ crystals~\cite{Pattavina_2021}.

In the present work, we exploit the properties of archaeological Pb to develop and operate a cryogenic detector based on PbWO$_4$ crystals. This approach allows us to investigate the potential of this material for direct DM detection, while simultaneously exploring its suitability for CE$\nu$NS searches. The combination of a high-$A$ target, a low energy threshold, and an ultra-low intrinsic background represents a promising pathway toward the observation of rare nuclear recoils associated with physics beyond the Standard Model~\cite{Drukier:1983gj}.

A preliminary study of the sensitivity of PbWO$_4$ to both spin-dependent and spin-independent DM interactions was presented in Ref.~\cite{RES-NOVACollaboration:DM2025}. In that work, the RES-NOVA collaboration exploits the presence of both heavy (Pb) and light (O) nuclei to explore a wide range of DM masses spanning nearly four orders of magnitude. With an expected target mass of approximately 200~kg and a total exposure of 1~yr, RES-NOVA is projected to probe previously unexplored regions of the DM parameter space, providing complementary information to that obtained by established technologies such as Xe-based time-projection chambers~\cite{XENON:2023cxc,PandaX_2021,LZ:2022lsv}.

\begin{table}
\caption{Evaluated internal radioactive contaminations for the PbWO$_4$ crystal produced from archaeological Pb. Nuclides of the decay chains with longer half-lives are listed. Limits are at 90\% C.L. \cite{kg-scale}} 
\begin{center}
\begin{tabular}{lcc}
\hline\noalign{\smallskip}
Chain & Nuclide  & Activity \\ 
            & & [mBq/kg] \\
\noalign{\smallskip}\hline\noalign{\smallskip}
$^{232}$Th & $^{232}$Th & $<$0.04 \\
 & $^{228}$Th & 0.80$\pm$0.09 \\
\noalign{\smallskip}\hline\noalign{\smallskip}
$^{238}$U & $^{238}$U & $<$0.03 \\
& $^{234}$U & $<$0.03 \\
& $^{230}$Th & $<$0.04 \\
& $^{226}$Ra & 11.34$\pm$0.35 \\
& $^{210}$Pb/$^{210}$Po & 22.50$\pm$0.49 \\
\noalign{\smallskip}\hline\noalign{\smallskip}
$^{40}$K & $^{40}$K & $<$9 \\
\noalign{\smallskip}\hline
\end{tabular}
\label{tab:radio} 

\end{center}
\end{table}

\section{Experimental set-up}

The crystal used in this work was obtained by cutting a 13~g sample from the same kg-scale PbWO$_4$ crystal previously operated by the RES--NOVA collaboration in Ref.~\cite{kg-scale}. That large-mass crystal was specifically employed to assess the intrinsic radioactive background of the material, benefiting from the long accumulated exposure and the ultra--low--background conditions of the experimental setup. 

In particular, the kg-scale crystal was operated within the cryogenic infrastructure that previously hosted the CUORICINO~\cite{CUORICINO:2008jjc}, CUORE--0~\cite{Alfonso:2015wka}, and CUPID--0~\cite{Azzolini:2018tum} experiments, enabling a detailed characterization of its intrinsic radionuclide content. A summary of the measured radioactive contaminants is reported in Table~\ref{tab:radio}.

The detector was operated in the Ieti dilution refrigerator~\cite{IETI}, a custom-built cryostat designed for detector R\&D in rare-event searches and optimized for mechanical stability and vibrational isolation. The Ieti cryostat provides a dry dilution refrigeration cycle, reaching a base temperature below 7~mK and offering a cooling power of approximately 500~$\mu$W at 100~mK. Its compact design and reduced mechanical footprint make it ideal for both R\&D and small-scale demonstrator experiments, despite the relatively high-background operating conditions. This system is hosted in the Hall-C of the underground laboratory of Gran Sasso of INFN (Italy).

\subsection{Cryostat Configuration}
In cryogenic rare-event searches (e.g. DM searches) minimizing environmental and mechanical noise is paramount to reaching sub-keV energy thresholds. Low-frequency mechanical vibrations—primarily induced by the operation of the cryostat pulse tube refrigerator (PTR) can propagate through the thermal stages and couple to the sensitive thermal sensors (e.g. NTD thermistors) generating parasitic heat loads and microphonic noise. This mechanical noise directly degrades the detector's baseline energy resolution~\cite{ALDUINO20199} affecting the detector performance at low energies. Consequently, an extensive characterization and decoupling of the cryogenic vibrational environment is essential to validating the overall detector performance and noise mitigation strategy.

\begin{figure}
    \centering
    \includegraphics[width=0.8\linewidth]{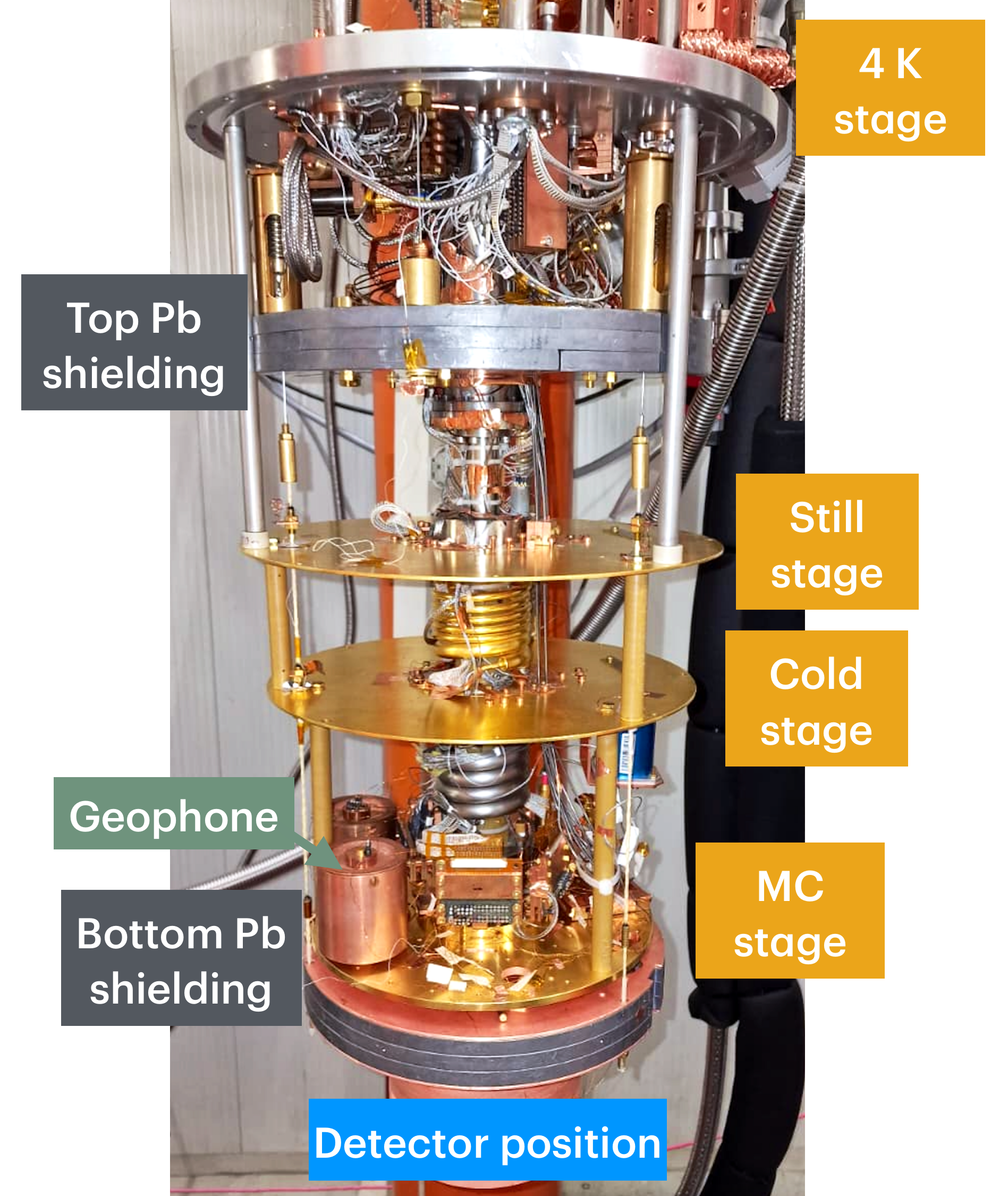}
    \caption{Photograph of the Ieti cryogenic infrastructure used for the PbWO$_4$ detector operation. The internal structure of the dilution refrigerator is shown, with the main temperature stages labeled (4~K stage, Still stage, Cold stage, and mixing chamber (MC) stage). The detector is installed at the MC stage, corresponding to the lowest temperature point of the cryostat. Passive Pb shieldings are placed above the detector to mitigate environmental $\gamma$-ray backgrounds while maintaining compatibility with the cryogenic and mechanical constraints of the setup. The installation position of the geophones is also indicated.}
    \label{fig:IETI}
\end{figure}

The cryostat has multiple thermal stages (50~K, 4~K, Still, Cold Plate, and Mixing Chamber) and a PTR for precooling, see Fig.~\ref{fig:IETI}. Each stage is thermally anchored to minimize temperature gradients and parasitic heat loads. The detector module is suspended from the Mixing Chamber (MC) plate using a low thermal conductivity structure engineered to reduce the transmission of mechanical vibrations, from the cryogenic infrastructure to the detector, while ensuring efficient thermal contact.

To further suppress the mechanical noise induced by the PTR, the cold head was mechanically decoupled from the cryogenic infrastructure and mounted on an independent support structure, physically isolated from the rest of the system. This configuration significantly reduces the transfer of vibrational noise to the detector volume.

The vibrational performance of the system was characterized using specially designed geophones operated at around 7~mK. These sensors enabled, for the first time, an in-situ assessment of vibrational noise within the actual operating environment of the detector (see Fig.~\ref{fig:IETI}). Previous characterizations on different cryogenic systems~\cite{vib_lyon, vib_muc} reported in the literature rely exclusively on piezoelectric sensors operated at room temperature or near-to-room temperature, which do not provide a complete or reliable picture of vibrational noise propagation under actual experimental conditions. Indeed, the thermal contraction of a cryogenic system can reduce the overall dimensions of a system up to the cm--scale, impacting its intrinsic mechanical response~\cite{deWit:2018itv}. The use of cryogenic geophones in this set-up thus represents a significant advancement in understanding the impact of mechanical noise on cryogenic rare-event detectors. In this work, we do not present any characterization of the mechanical noise at room temperatures, as dedicated studies are ongoing.

\begin{figure*}[]
    \centering
    \includegraphics[width=0.8\linewidth]{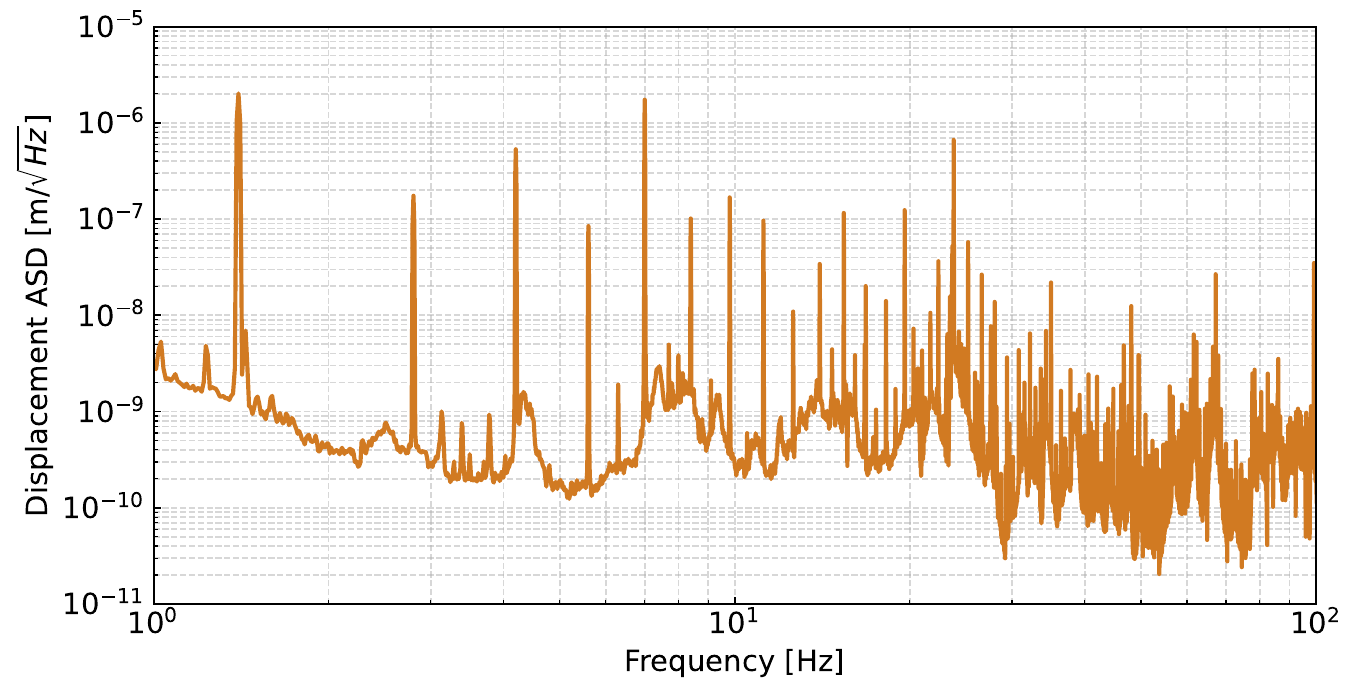}
    \caption{Ieti cryostat amplitude spectral density (ASD) of the axial displacement. This is measured at the Mixing Chamber plate during operation at 7~mK. The spectrum highlights the broadband vibrational background as well as narrow resonant features associated with mechanical modes of the cryogenic infrastructure.}
    \label{fig:vib}
\end{figure*}

In Fig.~\ref{fig:vib}, the axial (vertical) vibrational level of the Ieti MC plate is shown in terms of the displacement amplitude spectral density (ASD), measured at a temperature of 7~mK. When compared with vibrational measurements performed in other cryogenic experimental setups~\cite{vib_lyon, vib_muc}, the ASD indicates that the Ieti facility provides a particularly low--vibration environment, with vibrational amplitudes more than one order of magnitude lower in the low--frequency range (1--50~Hz), reaching the sub--nm level. This frequency range is particularly relevant for cryogenic detectors whose signal bandwidth covers this region.

\subsection{Detector set-up}
\begin{figure}
    \centering
    \includegraphics[width=0.95\linewidth]{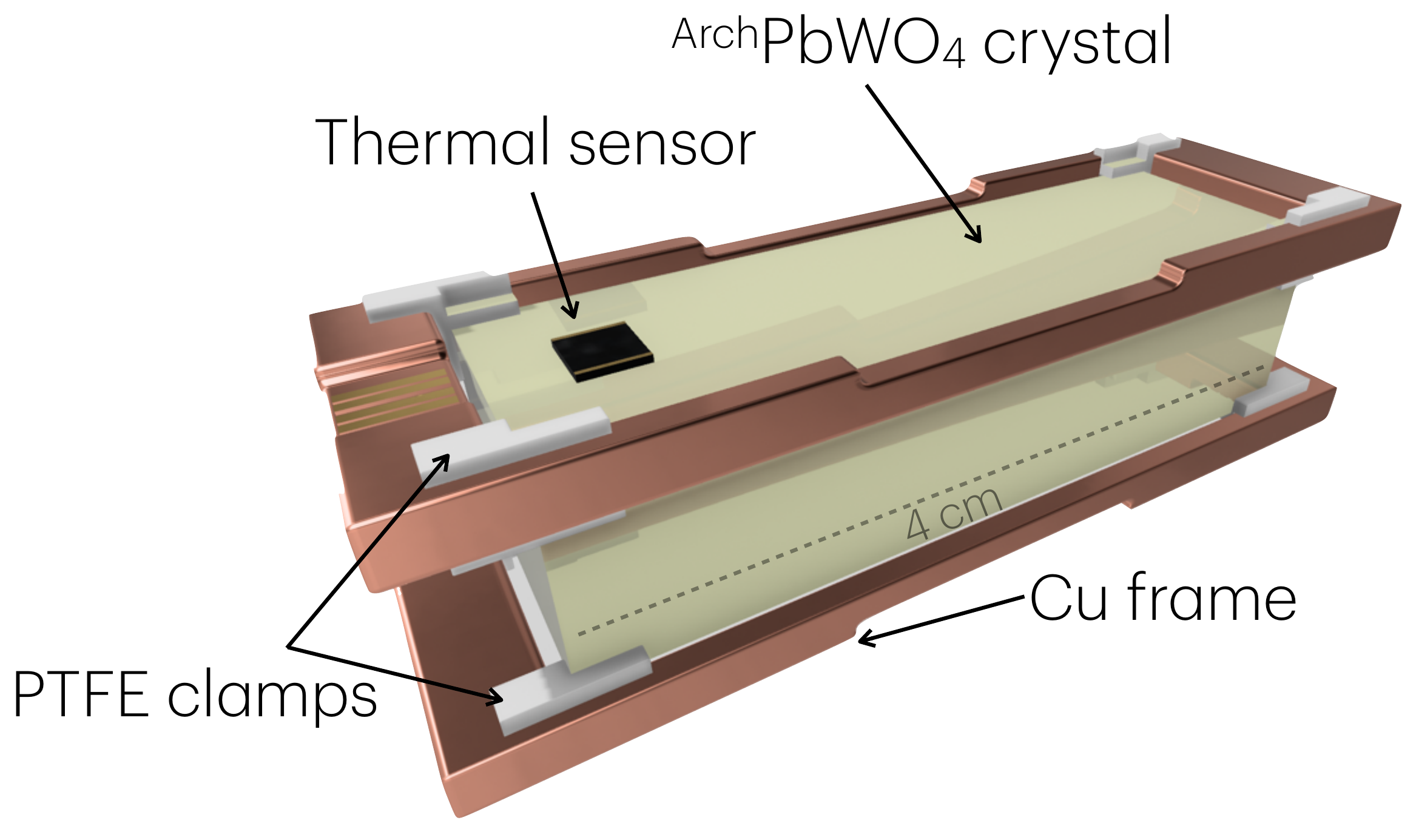}
    \caption{Schematic rendering of the cryogenic detector. The PbWO$_4$  crystal produced from archaeological Pb is held in position by PTFE clamps, which provide both mechanical support and a weak thermal coupling to the heat bath. The assembly is housed within a Cu frame that ensures mechanical stability and thermal anchoring to the cryogenic environment.}
    \label{fig:det}
\end{figure}

The detector consists of a PbWO$_4$ crystal shaped as a rectangular crystal with dimensions of 0.7~$\times$~0.7~$\times$~4~cm$^3$. The crystal is held in place and thermally coupled to the cryogenic environment by means of PTFE clamps, chosen for their low thermal conductivity and mechanical stability at mK temperatures. A Neutron Transmutation Doped (NTD) Ge thermistor~\cite{Pirro:2017ecr} is glued to the crystal using Araldite epoxy resin, providing sensitive thermometric read-out of the heat pulses induced by particle interactions.  
The crystal is mounted inside an oxygen-free high-conductivity (OFHC) copper frame. In Fig.~\ref{fig:det}, a schematic view of the PbWO$_4$ detector is shown.

The thermal signals from the NTD are read out via 25~$\mu$m diameter gold bonding wires, which provide the necessary electrical connection while maintaining minimal thermal load. These wires deliver the signals to a contact pad array located on the MC stage. From this point, superconducting wires are used to transport the signals outside the cryogenic environment to the room-temperature read-out electronics, thereby minimizing parasitic heat leaks and preserving signal integrity.

The entire setup was operated underground at the Laboratori Nazionali del Gran Sasso (LNGS), benefiting from an overburden equivalent to 3600~m.w.e.~\cite{Depth_LNGS}, which reduces the residual muon flux by approximately 6-orders of magnitude. The cryostat is housed within a passive shielding structure composed of non-archaeological Pb with a total thickness of 10~cm, surrounding the detector on the lateral sides.
The top side of the detector is shielded by an additional 6~cm of non-archaeological Pb, arranged in two separate disks of 3~cm thickness each and installed directly inside the cryostat, one at the 4~K stage and one below the MC stage (see Fig.~\ref{fig:IETI}). The lateral sides are further surrounded by a 5~cm-thick Cu shield.
This shielding configuration provides an effective suppression of environmental $\gamma$-ray backgrounds while maintaining mechanical compatibility with the cryogenic infrastructure. Nevertheless, residual background contributions are expected to originate from the shielding components themselves, which were not selected for radiopurity but rather optimized for shielding effectiveness and mechanical constraints.

\subsection{Readout and Electronics}
The thermistor was read out using a room-temperature setup. The bias was provided via a pair of high-value (1.3~G$\Omega$ + 1.3~G$\Omega$) load resistors characterized by low $1/f$ noise. These large resistance values minimize their parallel noise contribution. The maximum bias voltage across the load resistors is $50$~V.
This symmetric biasing circuit enables signal amplification through a differential voltage preamplifier—an improved version of the design in Ref.~\cite{Arnaboldi_2018}. This preamplifier features very low series noise, specifically below $4~\text{ nV}/\sqrt{\text{Hz}}$ at $1$~Hz and $1.4~\text{nV}/\sqrt{\text{Hz}}$ in the white noise region. Its parallel noise is approximately $0.3~\text{fA}/\sqrt{\text{Hz}}$, well below that of the load resistors. When accounting for the thermal noise of the detector’s connecting links, the total white series noise at the preamplifier stage is approximately $2 \text{ nV}/\sqrt{\text{Hz}}$.
The dedicated power supply system ensures high stability and negligible thermal drift for both front-end and detector biasing \cite{10.1063/1.4948390}. The preamplifier is followed by a programmable gain amplifier (8-bit resolution) to optimize the signal range for the acquisition system. This system \cite{CARNITI2023167658} consists of a frequency-programmable, 6-pole Bessel-Thomson antialiasing filter (ranging from $24$~Hz to $2.5$~kHz with 10-bit resolution) and a 24-bit $\Delta$-$\Sigma$ ADC. Data were acquired in continuous mode with a sampling frequency configurable up to $24$~kHz. The data stream was transmitted via a standard User Datagram Protocol (UDP) over an optical interface, enabling both online and offline processing.

For the measurements presented in this manuscript, the detector operating parameters were optimized to maximize the signal-to-noise ratio, following the same procedure described in Ref.~\cite{CUPID-0_det}. The bias voltage was set to 22~V (corresponding to a bias current of 8.5~nA), the amplifier gain to 1030, and the sampling frequency to 10~kHz. At base temperature (7~mK) and under operating conditions (12~mK), the Ge--NTD thermistor exhibited resistance values of 36~M$\Omega$ and 5.4~M$\Omega$, respectively.

\section{Data Processing and detector performance}

\begin{figure*}[]
  \centering
  \begin{subfigure}{0.5\textwidth}
    \centering
    \includegraphics[width=\linewidth]{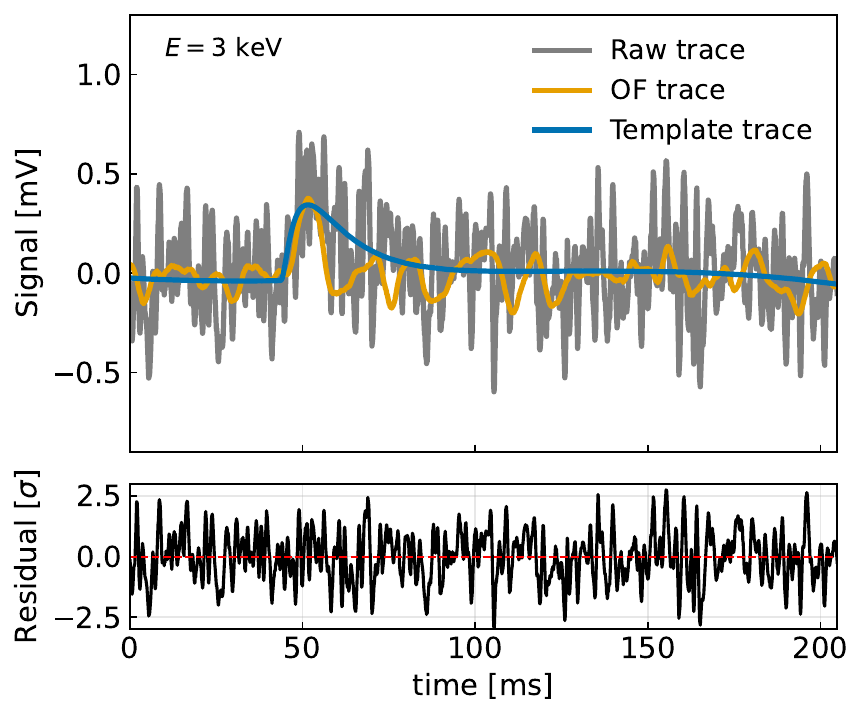}
    \label{fig:3keVPulse}
  \end{subfigure}\hfill
  \begin{subfigure}{0.5\textwidth}
    \centering
    \includegraphics[width=\linewidth]{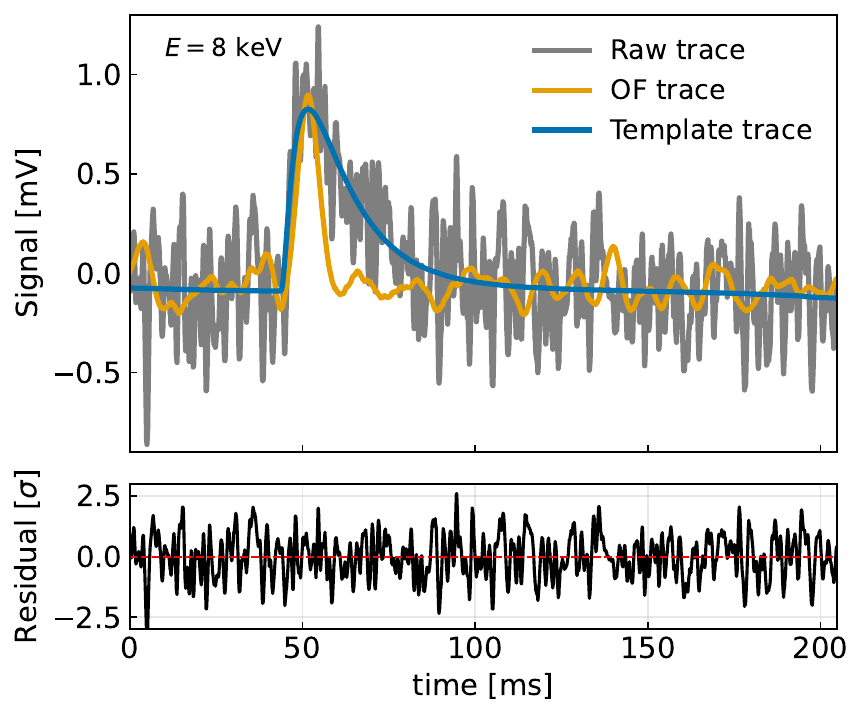}
    \label{fig:8keVPulse}
  \end{subfigure}
  \caption{Trace detector pulses at the end of the electronics read-out chain. The left (right) panel shows a typical 3~keV (8~keV) event. For each acquisition window, the raw trace (grey) is compared with the optimal-filter (OF) reconstructed trace (orange) and the corresponding signal template (dark blue). The lower panels show the normalized residuals between the raw trace and the best-fit template, expressed in units of the baseline noise standard deviation ($\sigma$). The residuals are centered around zero and do not exhibit significant time-dependent structures, demonstrating that the template accurately describes the pulse shape down to energies close to the analysis threshold (2.5~keV), where the signal-to-noise ratio approaches unity. }
  \label{fig:processed_pulses}
\end{figure*}

The continuous data stream, accounting for a total exposure of 32.4~g$\cdot$day, was processed through the newly developed analysis chain for RES--NOVA. The raw data stream was replicated into two parallel and synchronously processed branches. In the first branch, the signal was AC-coupled to remove the baseline offset and subsequently passed through a low-pass filter to suppress high-frequency noise. The filtered data were then processed in chunks of 2048 samples (204.8~ms), for a total of $\sim$780,000 chunks, which were treated as candidate pulse windows, as shown in see Fig.~\ref{fig:processed_pulses}. Each window was timestamped at the position of the maximum output of the filter, such that any particle-induced signal could be accurately located in the second, unfiltered data stream. Events extracted from this unfiltered branch were subsequently characterized using standard pulse-shape parameters.

This analysis pipeline operates in a trigger-less mode. When pulses are present, the filter output tags their position and amplitude; when no signal is present, the waveform is processed in the same way and timestamped at the filter's maximum output within the 2048 samples. This approach allows the Region of Interest (RoI) to be defined a posteriori, without imposing any threshold or selection bias at the data-acquisition stage.

In the first stage of the analysis, the identified pulses were used to characterize the detector response in terms of signal shape and noise power spectral density. The signal template was obtained by averaging approximately 1000 particle-induced pulses of similar amplitudes around 500~keV. The noise power spectral density was computed in an analogous manner by averaging the squared Fourier components of baseline segments with the same duration as the pulse windows but containing no physical events. These two ingredients were then combined to construct the Optimum Filter~\cite{Gatti:1986cw} (OF), which maximizes the signal-to-noise ratio under the assumption that the input waveform follows the expected signal shape and the stationary noise model.

In the second stage of the analysis pipeline, the OF replaced the generic low-pass filter, enabling optimized tagging of small pulses. In addition, pulses identified in the raw data stream—using the timestamps provided by the filtered branch—were fitted to a pulse model. The model consisted of an ideal signal template supplemented by a third-order polynomial to account for slow baseline drifts. Pulse amplitudes were extracted using a Maximum-Likelihood Estimation (MLE) procedure. In Fig.~\ref{fig:processed_pulses}, we show two representative pulse traces at different energies: one close to the detector energy threshold (at 3~keV) and one at higher energy (at 8~keV). The robustness of the pulse model is shown in the normalized residuals of Fig.~\ref{fig:processed_pulses}. The pulse near the detector energy threshold exhibits a signal--to--noise ratio close to unity.

This pipeline yields two independent estimators for the amplitude of particle-induced pulses: the OF amplitude and the MLE amplitude. The OF amplitude typically provides the best resolution, as it explicitly maximizes the signal-to-noise ratio under the assumption of a known pulse shape and stationary noise. The MLE amplitude, on the other hand, is obtained by fitting each individual pulse to an idealized detector-response model and is therefore more sensitive to pathological or poorly reconstructed events. In particular, since MLE relies on a constrained $\chi^2$ minimization, deviations from the expected pulse shape, such as pile-up, glitches, or baseline instabilities, can lead to biased or unstable amplitude estimates.

\begin{figure}[]
    \centering
    \includegraphics[width=1.0\linewidth]{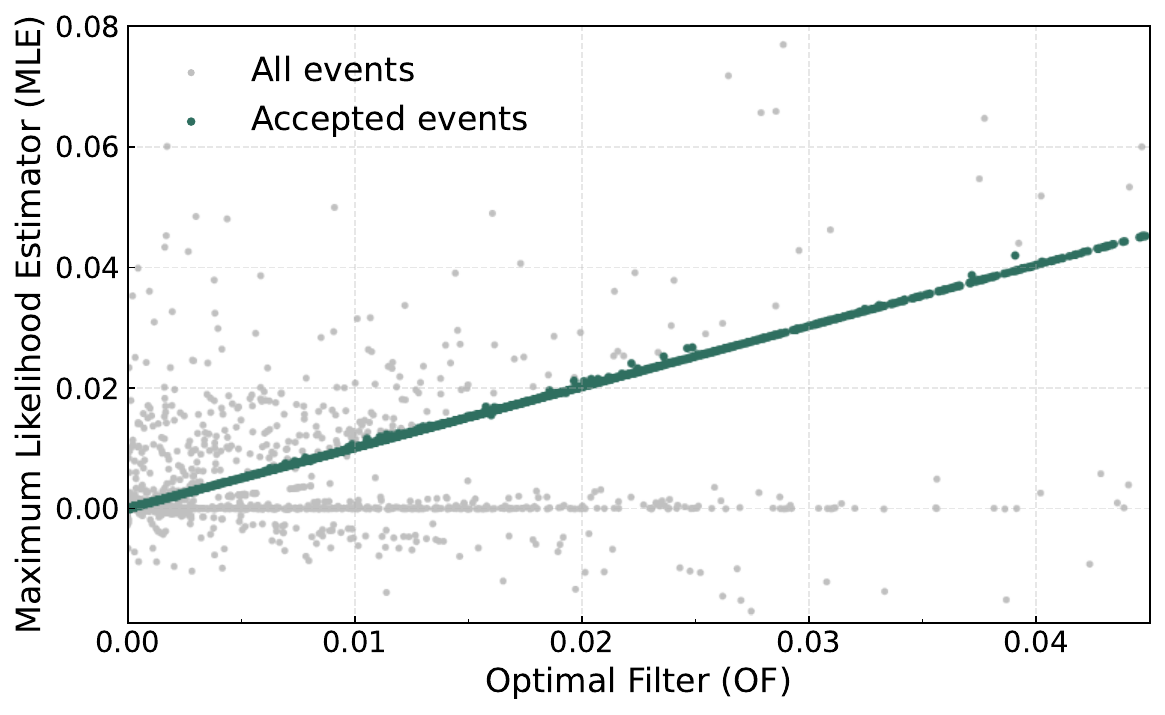}
    \caption{Comparison between the reconstructed pulse amplitudes obtained with the optimal filter (OF) and the maximum likelihood estimator (MLE). Each point corresponds to a single event. The full dataset is shown in light gray, while events passing the pulse-shape quality selection are highlighted in green. Accepted events cluster along the diagonal, indicating consistency between the two estimators, whereas rejected events populate regions with significant deviations from linearity. This behaviour reflects the effectiveness of the quality selection in removing noise-dominated or poorly reconstructed events from the analysis sample.}
    \label{fig:data_cut}
\end{figure}

A simple event–selection criterion was adopted: the amplitude obtained from the OF was required to be consistent, within a given tolerance, with the amplitude extracted from the pulse fit:
\begin{equation}
\frac{|A_{\mathrm{OF}}-A_{\mathrm{MLE}}|}{A_{\mathrm{OF}}} < \delta
\end{equation}
We deliberately avoid a direct $\chi^2$-based selection in order to minimize sensitivity to small template mismatches and to maintain a conservative event selection. This condition ensures that only well–reconstructed pulses, whose shapes are compatible with the expected detector response, are retained for the subsequent analysis. 
A comparison between the amplitudes reconstructed with the OF and the MLE is shown in Fig.~\ref{fig:data_cut}. Events passing the pulse-shape quality selection cluster along the diagonal, indicating a good agreement between the two estimators and their linear behavior, while rejected events populate off-diagonal regions. This demonstrates the effectiveness of the selection in identifying and removing poorly reconstructed or noise-dominated events.

Several tolerance values were tested independently, namely 5\%, 10\%, 20\%, and 30\%. Since this parameter strongly influences the final sample of accepted events, its variation provides an estimate of the systematic uncertainty associated with the selection procedure. For the analysis presented here we adopted a tolerance of 10\%, yielding 1,570 DM candidate events.
A summary of the main experimental and analysis parameters is presented in Tab.~\ref{tab:analysis_summary}.

The pulse amplitude ultimately used for the analysis is that provided by the OF, which offers the optimal energy resolution.

\begin{table}[t]
\caption{Summary of the main analysis parameters used for the Dark Matter search.}
\label{tab:analysis_summary}
\centering
\begin{tabular}{lc}
\hline
Quantity & Value \\
\hline
Detector mass & 13~g \\
Total exposure & 32.4~g\,d \\
Processed windows & $\sim 7.8\times10^{5}$ \\
Window length & 204.8~ms \\
Analysis threshold & 2.5~keV \\
RoI & 2.5--10~keV \\
OF--MLE tolerance & 10\% \\
DM candidate events & 1570 \\
Baseline resolution & $\sigma = 234$~eV \\
RoI background rate ($e^-/\beta$+NR) & $\sim 10^{3}$~c/keV/kg/d \\
\hline
\end{tabular}
\end{table}

\begin{figure*}[h]
  \centering
  \begin{subfigure}{0.5\textwidth}
    \centering
    \includegraphics[width=\linewidth]{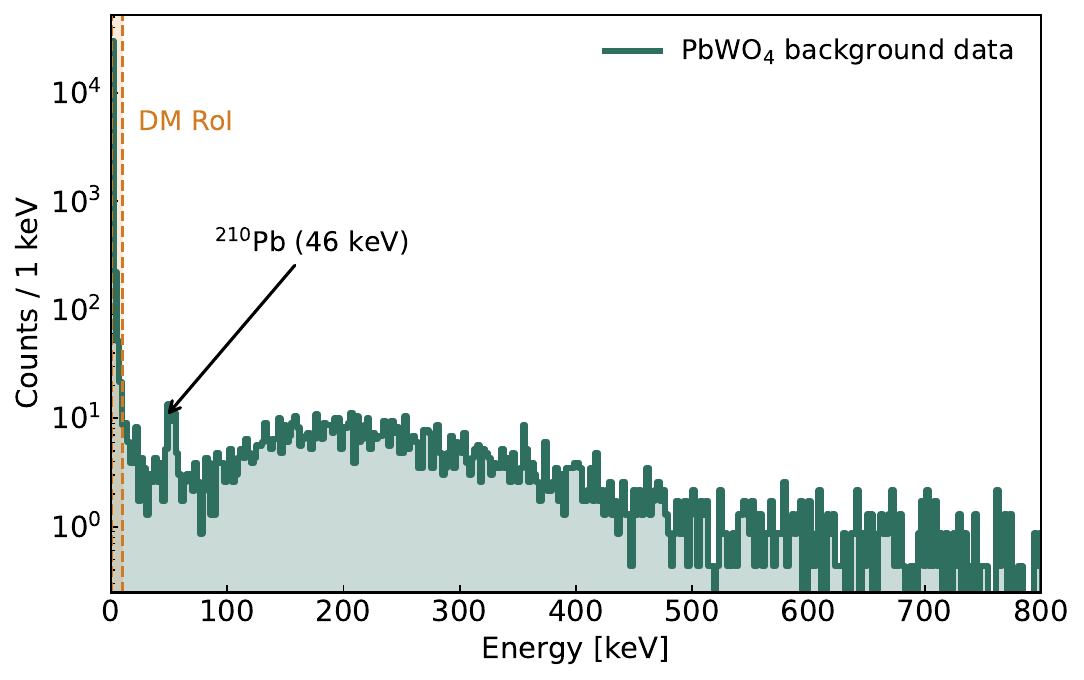}
  \end{subfigure}\hfill
  \begin{subfigure}{0.5\textwidth}
    \centering
    \includegraphics[width=\linewidth]{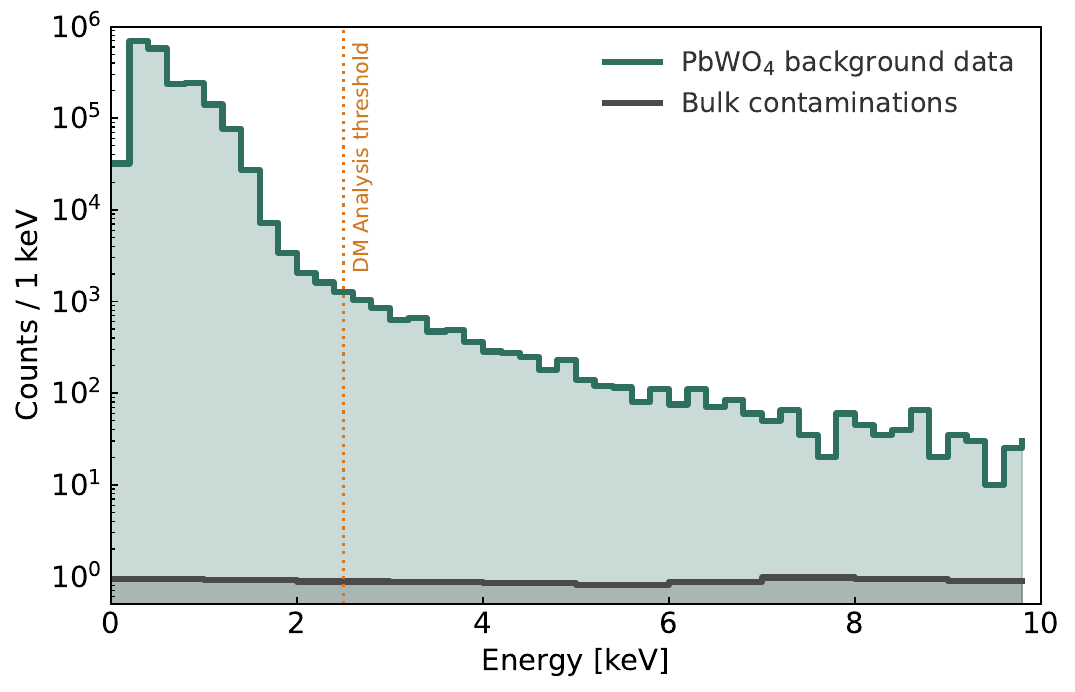}
  \end{subfigure}
  \caption{Background energy spectra measured of the PbWO$_4$ cryogenic detector. \textbf{Left:} energy spectrum of all accepted events, including $e^{-}/\gamma$ interactions and nuclear recoils, measured up to 800~keV. Apart from the peak at 46~keV, attributed to the characteristic emission of $^{210}$Pb, no other clear spectral features are observed. The region of interest for the DM analysis (2.5--10~keV) is highlighted in orange. Data acquired over an exposure of 32.4~g$\cdot$d are reported as a green histogram.
\textbf{Right:} zoom in up to 10~keV of the background energy spectrum. For comparison, the expected contribution from radioactive bulk contaminations of the crystal, derived from Ref.~\cite{kg-scale}, is shown as a grey histogram. The measured spectrum is dominated by external backgrounds originating from the cryogenic infrastructure (e.g.\ $^{210}$Bi bremsstrahlung from the internal Pb shielding) and from the surrounding environment (e.g.\ ambient $\gamma$ and neutron radiation).
}\label{fig:accepted_events}
\end{figure*}

The measured pulse amplitudes were converted into deposited energy by identifying the 2615~keV $\gamma$ line from $^{208}$Tl in the amplitude spectrum of a calibration measurement, which was used to anchor the absolute energy scale of the detector. As an independent validation of the calibration, we verified that the calibrated spectrum correctly reproduces the characteristic onset at 46~keV associated with the combined gamma and beta emissions from $^{210}$Pb, also in the background measurement (see Fig.~\ref{fig:accepted_events}). Thanks to these clear signatures in the energy spectrum, we can derive the detector signal amplitude: 112$\cdot 10^{-3}$~$\mu$V/keV, a value that is in line with other high-performance cryogenic detectors~\cite{Beeman:2012wz, Cardani:2011vg, Barucci:2019ghi, Helis:2024vhr}.

Given the absence of calibration points below 46~keV, the energy calibration in the RoI relies on an extrapolation that necessarily relies on an extrapolation below the lowest calibration point\footnote{Ge-NTDs provide linear energy response~\cite{Pirro:2017ecr} and this treatment is employed because of the lack of a validation for the energy calibration in the RoI. The final RES-NOVA implementation will rely on TES that will be tested with dedicated calibration lines at low energy.}. To quantify this effect, we consider a set of polynomial calibration models of increasing order, fitted to the available data with a free intercept, and evaluate their stability under the removal of high-energy calibration points. We have identified the following characteristics $\gamma$-peaks: 46~keV ($^{210}$Pb), 511~keV (annihilation), 583~keV ($^{208}$Tl), 1461~keV ($^{40}$K) and 2615~keV($^{208}$Tl). The non-linearity uncertainty is defined as the envelope of the resulting inverse calibration functions, expressed as a variation of the reconstructed true energy at fixed measured energy.
This procedure intentionally overestimates the possible non-linearity in the RoI, providing a conservative uncertainty estimate in the absence of dedicated low-energy calibration lines. In order to prevent non-physical extrapolations leading to artificially low effective thresholds, a lower bound of 0.5~keV is imposed on the energy-scale variation. This truncation ensures that the resulting uncertainty remains physically meaningful and conservative in the region of interest.

In Fig.~\ref{fig:accepted_events}, the background energy spectrum of all accepted events, including $e^{-}/\gamma$ interactions and nuclear recoils, is shown. The spectrum exhibits a steep rise at low energies, around $2~\mathrm{keV}$, which originates from the acceptance of events with a signal--to--noise ratio close to unity. These events may include contributions of non-particle origin or signals whose pulse shape is inconsistent with the detector response model. Stricter selection criteria suppress the fraction of such \textit{noise--dominated} events in the accepted sample, while looser criteria, conversely, lead to an increased acceptance of these events. At the same time, stricter criteria reduce the overall detection efficiency, whereas looser selections lead to a higher acceptance. This suppressing or enhancing effect is observable down to energies of about $2.5~\mathrm{keV}$; below this threshold, the impact of the selection criteria on the spectrum is no longer appreciable, indicating a signal--to--noise ratio below unity. The $2.5~\mathrm{keV}$ threshold for event identification is primarily determined by the signal sensitivity of the thermal sensors employed in this study, namely Ge--NTD thermistors. In comparison with other sensor technologies better suited for low--energy DM searches, such as transition edge sensors (TES)~\cite{Pirro:2017ecr}, significantly lower thresholds are expected.

The dominant contribution to the total counting rate in the DM RoI (2.5-10~keV) originates from background sources external to the crystal. These are mainly associated with the cryogenic infrastructure, such as bremsstrahlung radiation from $^{210}$Bi present in the internal Pb shielding, as well as environmental backgrounds, mainly due to ambient neutrons and $\gamma$-ray radiation, as previously discussed in Ref.~\cite{RES-NOVACollaboration:DM2025}. 

Normalizing the spectrum in Fig.~\ref{fig:accepted_events} to the total exposure of 0.032~kg$\cdot$d, the average background rate in the RoI corresponds to $\sim 10^3$~c/keV/kg/d after quality selection.
Order-of-magnitude estimates indicate that brems--strahlung of $^{210}$Bi and radioactivity in the shielding of the experimental set-up can contribute at the level of $\sim10^3$~c/keV/kg/d, while environmental $\gamma$ radiation, mainly arising from the $^{222}$Rn decay chain, contributes at the level of $\sim 10^1$~c/keV/kg/d. The contribution from ambient neutrons is expected to be smaller, at the level of $\sim 1$~c/keV/kg/d.

Dedicated Monte Carlo studies will be performed to assess in more detail the different contributions. To further illustrate this point, Fig.~\ref{fig:accepted_events} also shows the expected contribution from radioactive bulk contaminations within the operated PbWO$_4$ crystal. This component is evaluated through Monte Carlo simulations of the experimental setup, using as input the results of the high-statistics radioactive screening measurements performed on the same crystal and summarized in Tab.~\ref{tab:radio}. The main contribution in the RoI is attributed to $^{210}$Pb.

\begin{figure}[]
    \centering
    \includegraphics[width=1.0\linewidth]{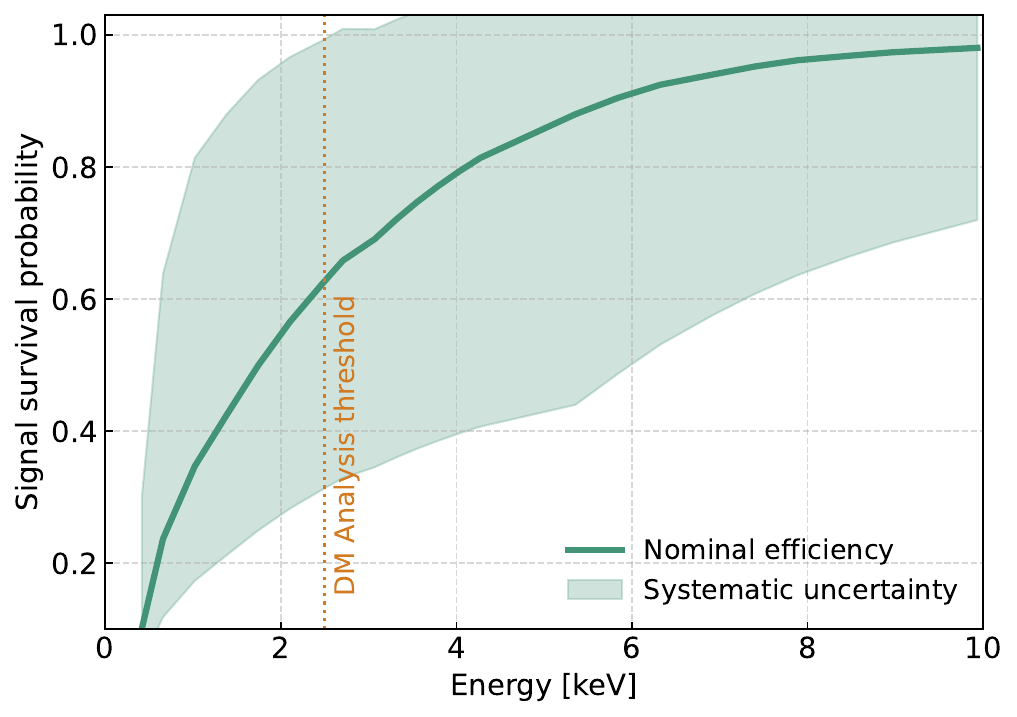}
    \caption{Signal survival probability as a function of the deposited energy. The curve represents the cumulative efficiency of the analysis cuts applied to simulated PbWO$_4$ events. The band represents the effect of the energy-calibration uncertainty on the signal survival probability.}
    \label{fig:efficiency}
\end{figure}

\begin{figure}[]
    \centering
    \includegraphics[width=1.0\linewidth]{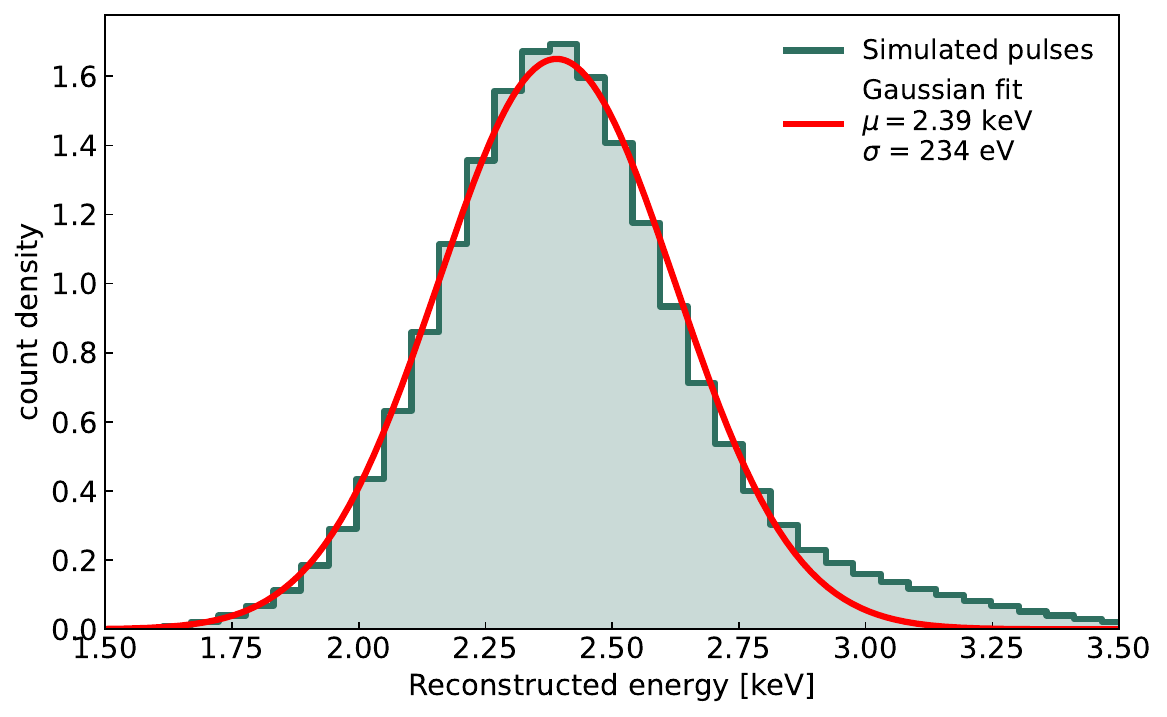}
    \caption{Optimum Filter energy reconstruction of a monochromatic line simulated at 2.4~keV, just below the DM analysis threshold ($\mathrm{S/N}\simeq 1$). The slight tail asymmetry is a consequence of using the maximum of the OF output as amplitude estimator.}
    \label{fig:baseline_resolution}
\end{figure}

To evaluate the detection and reconstruction efficiency in the RoI for DM investigations, we injected a comb of simulated mono-energetic pulses directly into the raw data stream and processed them through the full analysis pipeline. Injected pulses are generated as scaled copies of the signal template. No amplitude-dependent pulse-shape variations are included. This approximation is justified by the expected linear response of the Ge--NTD sensor in the considered energy range and by the observed linear correlation between pulse height and reconstructed amplitude down to the RoI.
Each data chunk of 2048 samples was tested for every simulated amplitude. For each simulated energy, events that passed the quality selection and were correctly reconstructed were counted as successful detections. The efficiency $\epsilon(E)$, dependent on the chosen quality selection tolerance, was then obtained as the fraction of successfully reconstructed events over the total number of injected pulses at each energy value. In Fig.~\ref{fig:efficiency}, the results of this analysis step are shown. In addition, for the evaluation of the detector baseline resolution, we fit a Gaussian distribution to the simulated lines obtaining: $\sigma=$~234~eV, see Fig.~\ref{fig:baseline_resolution}.


\section{Dark Matter search}
In order to derive the sensitivity to DM particles, we compare the observed energy spectrum with the expected one in PbWO$_4$ for a given DM particle mass and scattering cross-section. The expected recoil rate in the target crystal is computed using the \texttt{wimprates v0.5.0} package developed by the Xenon Collaboration~\cite{jelle_aalbers_2023_7636982}, including contributions from Pb, W, and O isotopes. For the spin-dependent interaction, we incorporate the nuclear spin structure of $^{207}$Pb and $^{17}$O, as discussed in Ref.~\cite{RES-NOVACollaboration:DM2025} and computed from Ref.~\cite{Bednyakov:2006ux}.

The parameters of the DM halo model are taken from Ref.~\cite{Baxter:2021pqo}, assuming a local DM density of $\rho_{\mathrm{DM}} = 0.3~\mathrm{GeV}/c^2/\mathrm{cm}^3$, an escape velocity of $v_{\mathrm{esc}} = 544~\mathrm{km/s}$ for the galaxy, a local standard-of-rest velocity of $v_{\mathrm{lab}} = 238~\mathrm{km/s}$, a solar peculiar velocity of $v_{\odot} = (11.1,\,12.2,\,7.3)~\mathrm{km/s}$, and an average galactocentric Earth speed of $v_{\oplus} = 29.8~\mathrm{km/s}$.  
The expected recoil spectrum, computed as a function of the true recoil energy $E_t$, is multiplied by the measured detection efficiency $\epsilon(E_t)$ and subsequently convolved with the detector response function, accounting for the finite energy resolution that maps $E_t$ into the reconstructed energy $E_r$. The resulting spectrum in reconstructed energy is then used as input for the Yellin optimum-interval calculation. In this analysis we assume a unity heat response for nuclear recoils, i.e. no quenching correction is applied to the reconstructed recoil energy, given that only a small fraction (1.8\%~\cite{Beeman:2012wz} of the deposited energy goes into scintillation. 

Although the experimental setup is installed underground, it is not optimized for ultra-low-background measurements. Moreover, Ge-NTD sensors do not provide the same performance as TESs for very small signal amplitudes, leading to a reduced efficiency of data selection at the lowest energies. As a consequence, no reliable background model can be assumed in the RoI. We therefore employ Yellin’s optimum interval method~\cite{Yellin:2002xd} to derive a robust—albeit conservative—estimate of the sensitivity to the DM particle–nucleon scattering cross-section at 90\%~CL.  

\begin{figure*}[h!]
  \centering
  \begin{subfigure}{0.5\textwidth}
    \centering
    \includegraphics[width=\linewidth]{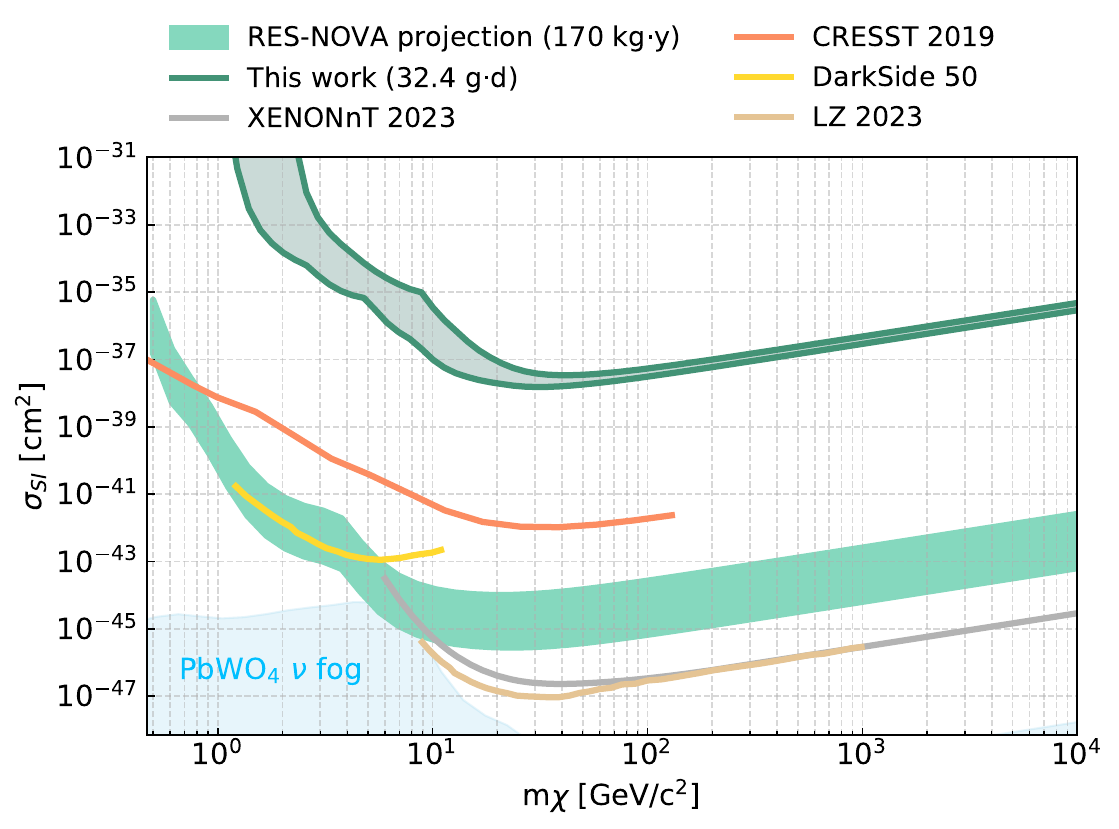}
  \end{subfigure}\hfill
  \begin{subfigure}{0.5\textwidth}
    \centering
    \includegraphics[width=\linewidth]{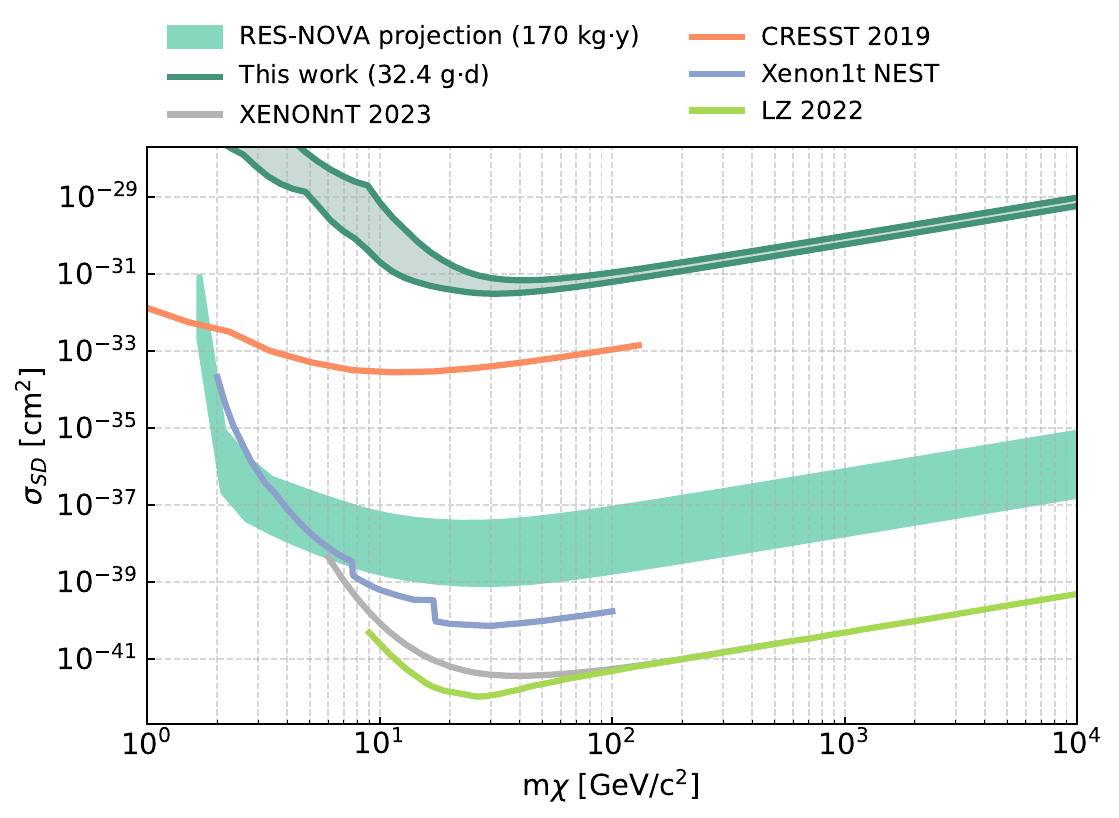}
  \end{subfigure}
  \caption{Upper limit on dark matter–nucleus scattering cross-section at 90\% confidence level. The left panel shows the limits on the spin-independent DM–nucleus scattering cross section as a function of the dark matter mass, while the right panel reports the corresponding constraints for spin-dependent interactions on $^{207}$Pb and $^{17}$O neutrons. The results obtained in this work, based on an exposure of 32.4~g$\cdot$d, are shown together with a RES--NOVA sensitivity projection assuming an exposure of 170~kg$\cdot$y. The band associated with the results of this work reflects the variation obtained by changing the pulse-shape consistency tolerance by 10\% and the uncertainty on the energy-calibration (dominant systematic effect). Existing limits from other direct detection experiments are also displayed for comparison~\cite{CRESST:2019jnq, DarkSide:2018bpj, XENON:2019gfn, LZ:2022lsv, XENON:2023cxc}. The present sensitivity is limited by the achieved exposure and by background contributions from the cryogenic infrastructure.}
  \label{fig:limits}
\end{figure*}

The lowest cross sections that can be excluded by this measurement, as a function of the DM particle mass, are shown in Fig.~\ref{fig:limits} for both spin--independent and spin--dependent interactions on $^{207}$Pb and $^{17}$O neutrons. In this evaluation, we chose not to consider $^{183}$ W (natural isotopic abundance of 14.3\%~\cite{IUPAP}, $J^{\pi}=1/2^-$~\cite{Kondev_2021}), as it has the same nuclear spin and parity as $^{207}$Pb but a lower natural isotopic abundance, making it a less sensitive nuclear target. An order-of-magnitude evaluation indicates that the contribution of $^{183}$W to the total spin-dependent event rate remains approximately 10\% over the DM mass range explored in this work.
These results represent the first constraints obtained using PbWO$_4$ operated as a target material. The achieved sensitivity is currently limited by the detector exposure, amounting to only $32.4~\mathrm{g\cdot d}$, and by the environmental background associated with the cryogenic infrastructure. Operation of the same detector in a lower--background environment would substantially extend the sensitivity of this approach in probing the DM parameter space.
In Fig.~\ref{fig:limits}, we also report the sensitivity projections for the operation of the full RES-NOVA detector demonstrator over an exposure of 170~kg$\cdot$y~\cite{RES-NOVACollaboration:DM2025}.

\section{Conclusions and perspectives}

We have successfully operated a cryogenic detector prototype based on a PbWO$_4$ crystal grown from archaeological Pb and established a robust data-analysis framework enabling the derivation of exclusion limits on DM interactions with ordinary matter, for the first time using this compound. Although the achieved sensitivity is not yet competitive with that of state-of-the-art experiments, this measurement provides a clear proof of concept of the detector technology. While not representative of the final RES--NOVA detector design, the prototype demonstrates effective control of mechanical noise in a cryogenic system at a level compatible with low-energy rare-event searches. Furthermore, the developed analysis methods — encompassing noise characterization, event reconstruction, and background handling — are readily transferable to Supernova neutrino detection via CE$\nu$NS. Together, these results represent a key validation step toward the scientific objectives of the RES--NOVA program.

The results presented in this work validate the use of PbWO$_4$ crystals grown from archaeological materials in cryogenic rare-event searches. A significant improvement in sensitivity is expected from future implementations employing optimized low-background experimental setups and TESs in place of Ge-NTD thermistors, which are known to provide superior performance at low energies. In combination, these developments are expected to substantially lower the achievable energy threshold and enhance background discrimination capabilities.

In addition, we have validated the full data-analysis pipeline, which is capable of processing the detector data stream in real time. In the context of dark matter searches, this live processing enables continuous monitoring of detector performance and will be instrumental for operating-point stabilization and quality control once TES-based readout is implemented. In the context of Supernova neutrino detection, the same pipeline provides the capability to identify collective low-energy excesses and to issue alerts within seconds of a burst, thereby enabling prompt multi-messenger follow-up observations.

\begin{acknowledgements}
The RES--NOVA collaboration thanks the directors and staff of the Laboratori Nazionali del Gran Sasso for their support, as well as the technical staff involved in the experiment. This work received funding from the European Union’s Horizon Europe programme through the ERC grant ERC--101087295 RES--NOVA. Additional financial support was provided under the National Recovery and Resilience Plan (NRRP) by the Italian Ministry of University and Research (MUR), funded by the European Union – NextGenerationEU, within the project “Advanced techniques for a next--generation isotopically enriched cryogenic Dark Matter experiment” (2022L2AXP2).
This work made use of the \texttt{Arby} software for \texttt{Geant4}--based Monte Carlo simulations, developed within the framework of the Milano--Bicocca activities and currently maintained by O.~Cremonesi and S.~Pozzi.
We gratefully acknowledge the University of Milano--Bicocca and INFN for their continuous support of the collaboration. S.~Ghislandi acknowledges support from the U.S. Department of Energy (DOE) Grant No. [DE-SC0011091]. F.A.~Danevich and V.I.~Tretyak thank the Laboratori Nazionali del Gran Sasso, the Gran Sasso Science Institute, and the RES--NOVA Collaboration for their outstanding support and hospitality during the difficult period following the Russian invasion of Ukraine.

\end{acknowledgements}
\newcommand*{\doi}[1]{\href{https://doi.org/\detokenize{#1}}{DOI: \detokenize{#1}}}

\bibliographystyle{spphys}       
\bibliography{biblio}   

\end{document}